\documentclass[12pt]{article}

\usepackage{scicite}

\usepackage{times}

\usepackage{graphicx}

\newenvironment{sciabstract}{%
\begin{quote} \bf}
{\end{quote}}

\newcounter{lastnote}
\newenvironment{scilastnote}{%
\setcounter{lastnote}{\value{enumiv}}%
\addtocounter{lastnote}{+1}%
\begin{list}%
{\arabic{lastnote}.}
{\setlength{\leftmargin}{.22in}}
{\setlength{\labelsep}{.5em}}}
{\end{list}}

\title{Comment on ``Correlation of the Highest-Energy Cosmic Rays with Nearby
Extragalactic Objects''}

\author
{Dmitry Gorbunov,$^{1}$ Peter Tinyakov,$^{1,2}$ Igor Tkachev,$^{1}$
  Sergey Troitsky$^{1\ast}$\\
\\
\normalsize{$^{1}$Institute for Nuclear Research of the Russian Academy of
Sciences,}\\
\normalsize{60th October Anniversary Prospect 7a, 117312, Moscow, Russia}\\
\normalsize{$^{2}$Service de Physique
Th\'eorique, Universit\'e Libre de Bruxelles,}\\
\normalsize{ CP225, blv.~du Triomphe, B-1050, Bruxelles, Belgium}\\
\\
\normalsize{$^\ast$
E-mail:  st@ms2.inr.ac.ru}
}

\date{}

\begin{document}


\baselineskip24pt


\maketitle


\sloppy

\begin{sciabstract}
We argue that the data published by the Pierre Auger Collaboration
(arXiv:0711.2256) disfavor at 99\%
confidence level their hypothesis that most of the highest-energy
cosmic rays are protons from nearby astrophysical sources, either
Active Galactic Nuclei or other objects with a similar spatial
distribution.
\end{sciabstract}

The Pierre Auger Collaboration reported a remarkable correlation
\cite{PA} between the arrival directions of ultra-high energy cosmic
rays (UHECR) and positions of nearby Active Galactic Nuclei (AGN). The
correlation was found by scanning over the maximum angular separation,
the minimum event energy and the maximum AGN redshift, see
Ref.~\cite{TT} for the details of the method. The best signal was
found at the angle of 3.1$^\circ$ for the cosmic-ray set consisting of
15 events with reconstructed energies $E>5.6\times 10^{19}$~eV and for
the set of 472 AGN obtained by imposing the cut on the redshift, $z
\le 0.018$, in the catalog~\cite{Veron}. The correlation was tested
with the independent set consisting of 13 events, with the parameters
fixed {\it a~priori} from the first data set. The probability that the
correlation has occurred by chance is $1.7\times 10^{-3}$ as derived
from the independent set. The conclusion was made that the
anisotropy of arrival directions is consistent with the hypothesis
that ``most of the cosmic rays reaching Earth in that energy range are
protons from nearby astrophysical sources, either AGN or other objects
with a similar spatial distribution.'' We refer to this proposition as the
``AGN hypothesis'' in what follows. Crucial ingredients of this hypothesis
are nearly rectilinear propagation of UHECR and a
large~\cite{footenoteDDT} number of sources distributed similarly to AGN
which in turn follow the matter distribution~\cite{AGN-distrib}.

In this Comment we would like to point out that, given the data
presented in Ref.~\cite{PA}, the AGN hypothesis is unlikely. We should
stress that we question neither the fact nor the derived significance
of the correlation. It is the interpretation of Ref.\cite{PA} that we
put in doubt.

The flux of a given source decreases as $1/r^2$ with the distance $r$
to the observer. This is not taken into account in the method of
positional correlations used in Ref.~\cite{PA}. Here we
present statistical tests which include this suppression factor.

The distribution of matter (and, therefore, of AGN) in the nearby
Universe is very inhomogeneous. The role of local inhomogeneities is
enhanced by the cosmic-ray attenuation that cuts off the (uniform)
flux coming from distant sources. The AGN hypothesis implies that
when the suppression factors are properly taken into account, major
local structures such as the Centaurus and Virgo superclusters
provide sizeable contributions to the cosmic-ray flux at highest energies.

This prediction of the AGN hypothesis allows for statistical tests
different from the small-scale correlation analysis.  Fig.~1
shows the UHECR events used in the analysis of Ref.\cite{PA} together
with the expected flux of cosmic rays
simulated assuming the AGN hypothesis. One may identify the
Virgo and Centaurus superclusters. The expected numbers of events from AGN
in these two structures are nearly equal. It is seen in Fig.~1 that there
is a deficit of observed events from Virgo as well as from other local
structures, except the Centaurus supercluster. This suggests that the AGN
hypothesis proposed in Ref.\cite{PA} may be disfavored.

To quantify this statement we took the sample of AGN used in the
analysis of Ref.~\cite{PA}. According to the catalog
classification~\cite{Veron}, this sample consists of 457 AGNs, 14
quasars and 1 probable BL Lac object, Cen~A. We removed from the
sample 3 objects with $z=0$ classified as stars in the database
\cite{NED}. We calculated the expected number of events within given
angular distance from the center of the Virgo cluster assuming $1/r^2$
suppression of the flux, and compared it to the data. The results are
presented in Fig.~2. The observed and expected distributions of events
are inconsistent.  According to the Kuiper test, the probability that
the observed and simulated events are drawn from the same distribution
is $2\times 10^{-4}$. The main origin of inconsistency is clear: of 27
events, $\sim 6$ are expected to come from Virgo under the AGN
hypothesis while none is observed. The probability of this is
$10^{-3}$, in agreement with the Kuiper test.

Similar results are obtained in tests which do not use the Virgo
supercluster as a reference point. Comparing the Galactic longitudes of
observed and expected cosmic rays we find that the probability that
the two samples are drawn from the same distribution is 2\% according
to the Kuiper test, while for the Galactic latitude the corresponding
probability is $10^{-4}$. Analogous tests for the Supergalactic
longitude and latitude give the probabilities of $7\%$ and $10^{-4}$,
respectively. We conclude that the AGN hypothesis of Ref.~\cite{PA} is
disfavored at the confidence level of at least~99\%.

If, as it is suggested by the above arguments, the highest-energy
cosmic rays observed by the Pierre Auger Observatory do not come from
sources that follow the local matter distribution, how can one explain
the observed correlations with AGN? One possible explanation could be
the existence of a nearby bright source which happened to be in the
direction to the Centaurus supercluster where the density of
background AGN is larger than in average. Cen~A is a natural
candidate. Contrary to the AGN
hypothesis, this explanation would imply one or at most a few sources
in the nearby Universe and large deflections, due to either strong
magnetic fields or to the presence of heavy nuclei in the cosmic-ray
flux as suggested in Refs.~\cite{PAO-comp,Yak-comp}. In that case the
properties of the highest-energy cosmic rays may appear different if seen
from the Southern and Northern hemispheres.

To summarize, the data presented in Ref.\cite{PA} disfavor the
hypothesis that most of the highest-energy cosmic rays come from
nearby astrophysical sources, either AGN or other objects with a
similar spatial distribution. The alternative explanation of the
observed correlations could be, e.g., the existence of a bright source
in the direction of the Centaurus supercluster, Cen A being a possible
candidate.

\begin{figure}[p]
\begin{center}
\includegraphics [width=0.8\textwidth]{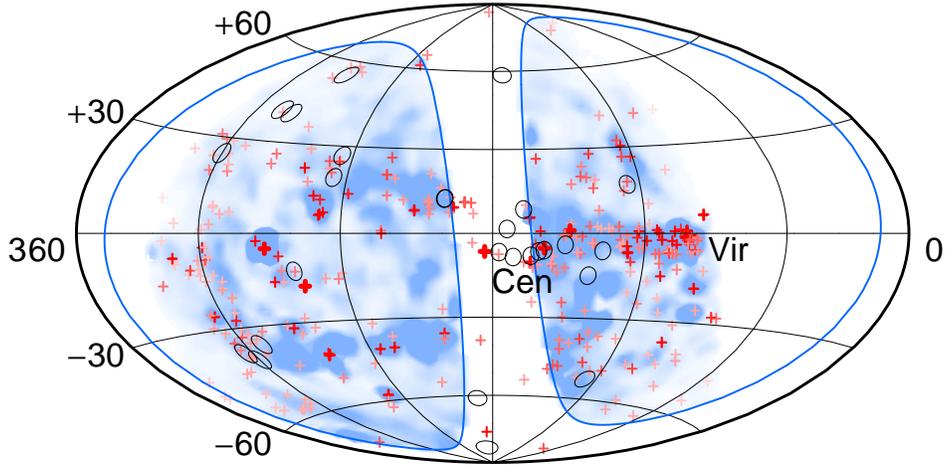}
\end{center}
\caption{Hammer projection of the celestial sphere in supergalactic
coordinates with crosses at the positions of nearby AGN from the
sample used in the correlation analysis of Ref.~\cite{PA}. The
color saturation of a given cross indicates the expected cosmic-ray flux
with the account of the acceptance of the Pierre Auger Observatory (PAO)
and the $1/r^2$ suppression, $r$ being the distance to the source. Open
circles represent 27 highest-energy cosmic rays detected by PAO. Shading
shows the expected cosmic-ray flux from sources that follow the local
matter distribution (for details see Ref.~\cite{khrenov}), smoothed at the
angular scale of $3.1^\circ$ and convoluted with the PAO acceptance
(darker regions correspond to larger cosmic ray flux). Blue lines cut out
the region with Galactic latitude $|b|< 15^\circ$ where the latter flux
cannot be determined because of incompleteness of the source catalog. The
positions of the Centaurus (Cen) and Virgo (Vir) superclusters are
indicated. }
\end{figure}

\begin{figure}[p]
\begin{center}
\includegraphics [width=0.8\textwidth]{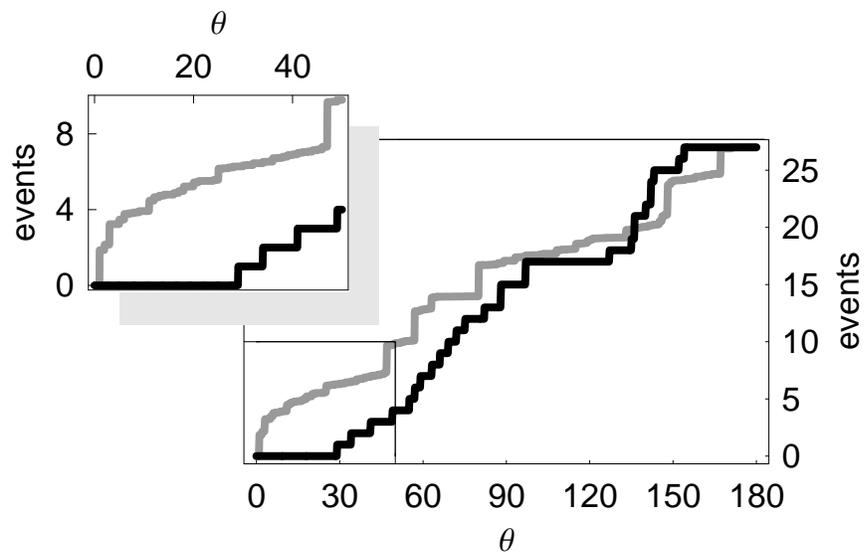}
\end{center}
\caption{
Number of events in the circle of radius $\theta$ (in degrees) centered on
the Virgo cluster as determined in~\cite{NED}: gray, expected number of
events assuming the AGN hypothesis; black, events actually observed,
Ref.~\cite{PA}. The side panel zooms on the region around Virgo.}
\end{figure}

\begin{scilastnote}
\item
We are indebted to Valery Rubakov and Mikhail Shaposhnikov for valuable
comments. This work was supported in part
by the grants NS-7293.2006.2
(government contract 02.445.11.7370) and RFBR 07-02-00820 (DG, IT and
ST), by the Russian Science Support Foundation (ST) and by the
Belspo:IAP-VI/11, IISN and FNRS grants (PT). ST thanks CERN and ULB for
hospitality. Numerical simulations have been performed at the computer
cluster of the Theoretical Division of INR RAS. This work has made use of
the NASA/IPAC Extragalactic Database~\cite{NED} operated by the Jet
Propulsion Laboratory, California Institute of Technology, under contract
with NASA.
\end{scilastnote}


\begin{thebibliography}{11}
\bibitem{footenoteDDT}
According to the
statistics of cosmic-ray clustering at small angles~\cite{DTT}, the number
of sources contributing to the flux is expected to be at least of order
several hundred assuming the AGN hypothesis.
\end{thebibliography}
\end{document}